\documentclass[pra,aps,twocolumn,superscriptaddress,showpacs]{revtex4}

\usepackage{graphicx}

\usepackage{amsmath}
\usepackage{amsfonts}

\newcommand{\abs}[1]{\left|#1\right|}

\def\halffigwidth{7cm}

\newcommand{\rmi}{{\rm i}}

\newcommand{\ket}[1]{|#1\rangle}
\newcommand{\beq}{\begin{equation}}
\newcommand{\eeq}{\end{equation}}

\begin{document}
\title{Numerical Analysis of Coherent Many-Body Currents in a Single Atom Transistor}
\author{A. J. Daley}
\affiliation{Institute for Quantum Optics and Quantum Information of the Austrian Academy of Sciences, A-6020
Innsbruck, Austria} \affiliation{Institute for Theoretical Physics, University of Innsbruck, A-6020
Innsbruck, Austria}
\author{S. R. Clark}
\author{D. Jaksch}
\affiliation{Clarendon Laboratory, University of Oxford, Parks Road, Oxford OX1 3PU,
U.K.}
\author{P. Zoller}
\affiliation{Institute for Quantum Optics and Quantum Information of the Austrian Academy
of Sciences, A-6020 Innsbruck, Austria} \affiliation{Institute for Theoretical Physics,
University of Innsbruck, A-6020 Innsbruck, Austria}
\date{June 29, 2005}

\begin{abstract}
We study the dynamics of many atoms in the recently proposed Single Atom Transistor setup [A. Micheli, A. J.
Daley, D. Jaksch, and P. Zoller, Phys. Rev. Lett. \textbf{93}, 140408 (2004)] using recently developed
numerical methods. In this setup, a localised spin 1/2 impurity is used to switch the transport of atoms in a
1D optical lattice: in one state the impurity is transparent to probe atoms, but in the other acts as a
single atom mirror.  We calculate time-dependent currents for bosons passing the impurity atom, and find
interesting many body effects. These include substantially different transport properties for bosons in the
strongly interacting (Tonks) regime when compared with fermions, and an unexpected decrease in the current
when weakly interacting probe atoms are initially accelerated to a non-zero mean momentum. We also provide
more insight into the application of our numerical methods to this system, and discuss open questions about
the currents approached by the system on long timescales.
\end{abstract}

\maketitle

\section{Introduction}
\label{satvidal:secintro}
The recently proposed Single Atom Transistor (SAT) setup \cite{sat} provides new
opportunities to experimentally examine the coupling of a spin-1/2 system with bosonic and fermionic modes.
Such couplings form fundamental building blocks in several areas of physics. For example, atoms passing
through a cavity can allow the quantum non-demolition (QND) readout of single-photon states in quantum optics
\cite{cqnd}, and in solid state physics, such systems occur in Single Electron Transistors \cite{set}, in
studies of electron counting statistics \cite{levitov} and in the transport of electrons past impurities such
as quantum dots \cite{cazmas}.

\begin{figure}
\begin{center}
\includegraphics[width=6.5cm]{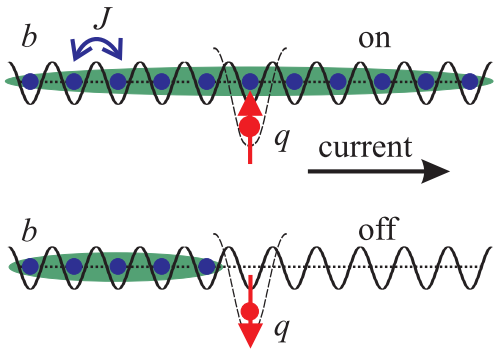}
\caption{A Single Atom Transistor (SAT) in a 1D optical lattice: A single spin-1/2 impurity atom $q$
separately trapped at a particular lattice site is transparent to probe atoms $b$ in one state (``on''), but
in the other acts as a single atom mirror (``off''). The probe atoms can either diffuse past the impurity
site with mean initial momentum $\langle \hat{k}\rangle_{t=0}=0$ or can be accelerated to a finite initial
momentum $\langle \hat{k}\rangle_{t=0} \neq 0$ by a kick of strength $p_k$.}\label{Fig:setup}
\end{center}
\end{figure}

In the SAT setup, which was motivated by the significant experimental advances made recently with cold atoms
in 1D \cite{esslinger1d,bloch1d,porto1d}, a single spin-1/2 impurity atom, $q$, is used to switch the
transport of a gas of cold atoms in a 1D optical lattice (Fig.~\ref{Fig:setup}). The impurity atom, which can
encode a qubit on two internal spin states, is transparent to a gas of probe atoms in one spin state (the
``on'' state), but acts as a single atom mirror in the other (the ``off'' state), prohibiting transport via a
quantum interference mechanism (Fig.~\ref{Fig:setup}). Observation of probe atoms that are initially situated
to one side of the impurity, and which can constitute either a 1D degenerage Bose or Fermi gas, can then be
used as a QND measurement \cite{qnd} of the qubit state of the impurity atom
$\ket{\psi_q}=\alpha\ket{\!\!\uparrow}+\beta\ket{\!\!\downarrow}$ \cite{sat} (see Fig.~\ref{Fig:setup}).

The long coherence times associated with atoms in optical lattices allow many-body effects to contribute
coherently to the transport properties over longer timescales than is observed in other systems where bosonic
and fermionic modes couple to a spin 1/2 system. This produces novel physics in which the current of atoms
passing the impurity, especially in a regime of weak coupling between probe atoms and impurity, is sensitive
to interactions between the probe atoms \cite{sat}. These effects could be directly observed in experiments,
for example, via measurements of the density of probe atoms on each site of the impurity atom as a function
of time.

In this article we present a detailed numerical analysis of these currents, making use of recently developed
numerical methods \cite{vidal} to calculate the dynamics of the bosonic probe atoms by directly integrating
the many-body Schr\"{o}dinger equation in 1D on an adaptively truncated Hilbert space. When these currents
are compared to analytical calculations of transmission coefficients for single particles passing the
impurity atom and the related currents for a non-interacting 1D Fermi gas, significant interaction effects
are observed, as first discussed in Ref. \cite{sat}. Here we provide new insight into the time dependence of
these currents, and what conclusions can be drawn from our numerical results on different timescales. We then
calculate the initial currents for atoms at zero temperature diffusing past the impurity (where the initial
mean momentum of the 1D gas, $\langle \hat{k} \rangle_{t=0}=0$, with $\hat{k}$ is the operator corresponding
to the quasi-momentum in the lowest Bloch band and $t$ the time), and explore the effects observed for
different interaction strengths of bosonic probe atoms. We then also investigate the currents for fermions
and bosons when the probe atoms are initially kicked ($\langle\hat{k}\rangle_{t=0}\neq 0$). This study is
complementary to the analytical study of the SAT that is given in recent article by Micheli et
al.~\cite{satandi}.

In section \ref{sec:overview} we discuss the basic physics of the SAT, and give a summary of the dynamics
found in \cite{sat} for single particles and non-interacting fermions. Then we present in detail the
numerical techniques that we use to compute the exact time evolution of the many-body 1D system. The
time-dependence of the resulting currents is discussed in section \ref{sec:res}, followed by a presentation
of the values of the initial steady state currents, both in the diffusive ($\langle\hat{k}\rangle_{t=0}=0$)
and kicked ($\langle\hat{k}\rangle_{t=0}\neq 0$) regimes. The conclusions are then summarised in section
\ref{sec:summary}.

\section{Overview}
\label{sec:overview}

\subsection{The Single Atom Transistor}
\label{sec:overviewsat}

\subsubsection{The System}

\begin{figure}
\begin{center}
\includegraphics[width=7cm]{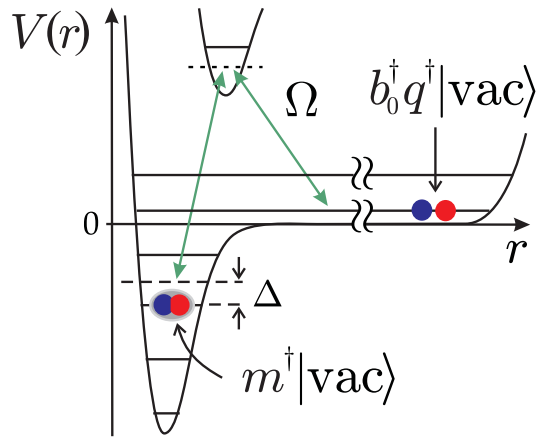}
\caption{An optical Feshbach resonance for a single spin channel ($\Omega=\Omega_\sigma$,
$\Delta=\Delta_\sigma$): One probe atom and the impurity atom, in an atomic state
$\hat{b}^\dag_0\hat{q}^\dag_\sigma \ket{{\rm vac}}$ which is quantised by the trapping potential of the
lattice site, are coupled by an optical Feshbach setup to a bound molecular state, $m^\dag_\sigma\ket{{\rm
vac}}$, of the Born-Oppenheimer potential, $V(r)$ (note that here the Born-Oppenheimer potential is modified
by the trapping potential of the lattice site \cite{juliennelattice}. The coupling has the effective
two-photon Rabi frequency $\Omega$, and detuning $\Delta$.}\label{Fig:feshbach}
\end{center}
\end{figure}

As described in section \ref{satvidal:secintro}, we consider probe atoms $b$, which are loaded into an
optical lattice \cite{hubbardtoolbox,greiner,oltheory,esslingerfermions} with strong confinement in two
dimensions, so that the atoms are restricted to move along a lattice in 1D. The probe atoms are initially
situated to the left of a site containing an impurity atom $q$, which is trapped independently (by a species
or spin-dependent \cite{sdol} potential), fixing it to a particular site while the probe atoms are free to
move. In order to produce the ``on'' and ``off'' states of the SAT, we must appropriately engineer the
effective spin-dependent interaction between the probe atoms and the impurity, $H_{\rm int}=\sum_\sigma
U_{\rm eff,\sigma}\hat{b}_0^\dag \hat{b}_0 \hat{q}_\sigma^\dag \hat{q}_\sigma$. Here, $\hat{b}_i^\dag$ and
$\hat{q}^\dag$ are second-quantised creation operators for the $b$ and $q$ atoms respectively, obeying the
standard commutation (anti-commutation) relations for bosons (fermions) and the site index $i$ is chosen so
that the impurity is on site $i=0$. These interactions can be controlled using either a magnetic
\cite{juliennelattice,magfeshbach} or optical \cite{optfeshbach} Feshbach resonance. For simplicity we
discuss the case of an optical Feshbach resonance, depicted in Fig.~\ref{Fig:feshbach}. Here, lasers are used
to drive a transition from the atomic state $\hat{b}^\dag_0 \hat{q}^\dag_\sigma \ket{{\rm vac}}$ via an
off-resonant excited molecular state to a bound molecular state back in the lowest electronic manifold
$\hat{m}^\dag_\sigma\ket{{\rm vac}}$ on the impurity site, $i=0$ (see Fig.~\ref{Fig:feshbach}). The
two-photon Rabi frequency for this process is denoted $\Omega_\sigma$ and the Raman detuning $\Delta_\sigma$,
and throughout this article we use units with $\hbar=1$.

\subsubsection{Single Atoms}

We consider initially a single probe atom passing the impurity. If the coupling to the molecular state is far
off resonance ($\Omega_\sigma \ll \abs{\Delta_\sigma}$), the effect of the Feshbach resonance is to modify
the interaction between the $b$ and $q$ atoms in the familiar manner, with $U_{\rm eff} =
U_{qb}+\Omega_\sigma^2/\Delta_\sigma$. This can be used to screen the background interaction between these
atoms, $U_{qb}$, so that the ``on'' state of the SAT ($U_{\rm eff}=0$) can be produced by choosing
$\Delta_\uparrow=-\Omega_\uparrow^2/U_{qb}$.

\begin{figure}
\begin{center}
\includegraphics[width=8cm]{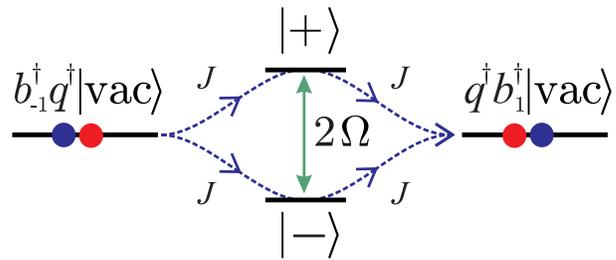}
\caption{The sequence as (left) a probe atom approaches the impurity site and is located on site $i=-1$,
(centre) the probe atom is on the impurity site, $i=0$, and (right) the probe atom has tunnelled past the
impurity and is located on site $i=1$. Quantum interference in this process because the two dressed states on
the impurity site, $\ket{\pm}=(\hat{b}^\dag_0\hat{q}^\dag_\sigma \ket{{\rm vac}} \pm m^\dag_\sigma\ket{{\rm
vac}})/\sqrt{2}$ give rise to two separate paths with equal and opposite amplitude. }\label{Fig:interfere}
\end{center}
\end{figure}

If the coupling is resonant ($\Delta_\downarrow=0$), then the physical mechanism is different, and the
passage of a probe atom $b$ past the impurity is blocked by quantum interference. The mixing of the unbound
atomic state and the molecular state on the impurity site produces two dressed states
\beq
\frac{1}{\sqrt{2}}\left(\hat{b}^\dag_0\hat{q}^\dag_\downarrow \ket{{\rm vac}} \pm m^\dag_\downarrow\ket{{\rm
vac}}\right),
\eeq
with energies
\beq
\varepsilon_\pm=\frac{U_{qb}}{2} \pm\left(\frac{U_{qb}^2}{4}+\Omega_\downarrow^2\right)^{1/2}.
\eeq
The two resulting paths for a particle of energy $\varepsilon$ then destructively interfere so that when
$\Omega_\downarrow \gg J$, where $J$ is the normal tunneling amplitude between neighbouring lattice sites,
and $U_{qb}=0$, the effective tunnelling amplitude past the impurity (see Fig.~\ref{Fig:interfere}) is
\beq
J_{\rm eff}=\left(-\frac{J^2}{\varepsilon + \Omega_\downarrow}-\frac{J^2}{\varepsilon -
\Omega_\downarrow}\right)\rightarrow 0.
\eeq
This is reminiscent of the interference effect which underlies Electromagnetically Induced Transparency
\cite{eit}, and corresponds to the effective interaction $U_{\rm eff}\rightarrow \infty$ required for the
``off'' state of the SAT.

In Refs.~\cite{sat,satandi}, the Lippmann-Schwinger equation is solved exactly for scattering from the
impurity of an atom $b$ with incident momentum $k>0$ in the lowest Bloch-band, where the energy of the
particle $\varepsilon(k)=-2J\cos(k a)$, with $a$ the lattice spacing. The resulting transmission
probabilities $T(p)$ are in the form of Fano Profiles \cite{satfano}. For $\Omega_\sigma \sim J$ these have a
minimum corresponding to complete reflection for $\varepsilon(k)=-\Delta_\sigma$ and complete transmission
for $\varepsilon(k)=-\Delta_\sigma-\Omega_\sigma^2/U_{qb}$. For $\Omega_\sigma>4 J$, the transmission
coefficients are approximately independent of $k$, and so complete transparency of the impurity atom is
obtained for $\Delta_\sigma=-\Omega_\sigma^2/U_{qb}$ and complete blocking of the incident atoms for
$\Delta=0$.

\subsubsection{Many Atoms}

The treatment of this system for many atoms is similar to the single atom case, but the motion of the probe
atoms in the lattice, except on the impurity site, is governed by a (Bose-) Hubbard Hamiltonian
\cite{oltheory}. As the two spin channels for the impurity atom, $q$ can be treated independently, we will
consider only a single spin channel $q_\sigma$, and drop the subscript in the notation throughout the
remainder of the article \cite{sat}. The Hamiltonian for the system is then given (with $\hbar\equiv 1$) by
$\hat{H}=\hat{H}_b+\hat{H}_0$, with
\begin{eqnarray}
\hat{H}_{\rm b}&=&-J\sum_{\langle ij \rangle} {\hat b}_i^\dag {\hat b}_j + \frac{1}{2} U_{bb} \sum_j
{\hat b}_j^{\dag}{\hat b}_j\left({\hat b}_j^{\dag}{\hat b}_j-1 \right), \nonumber \\
\hat{H}_{\rm 0}&=& \Omega \left({\hat m}^\dag {\hat q} {\hat b}_0 + {\rm h.c}\right) - \Delta {\hat m}^\dag
{\hat m}
\nonumber\\
& & + U_{qb} {\hat b}_0^\dag {\hat q}^\dag {\hat q} {\hat b}_0 + U_{bm} {\hat b}_0^\dag {\hat m}^\dag {\hat
m} {\hat b}_0.\label{Eq:Hamiltonian}
\end{eqnarray}
Here, $H_{\rm b}$ gives a Hubbard Hamiltonian for the $b$ atoms with tunnelling matrix elements $J$, and
collisional interactions $U_{bb}$. For fermions, $U_{bb}=0$, whereas for bosons $U_{bb}=4\pi\hbar^2
a_{bb}\int d^3\mathbf{x} \abs{{\rm w}_j(\mathbf{x})}^4/m_b$, with ${\rm w}_j(\mathbf{x})$ the
Wannier-function on site $j$, and $a_{bb}$ and $m_b$ the scattering length and mass of $b$ atoms
respectively. $H_0$ describes the dynamics in the presence of the impurity on site 0, where atoms $b$ and $q$
are converted to a molecular state with effective Rabi frequency $\Omega$ and detuning $\Delta$, and the
final two terms describe background interactions, $U_{\alpha\beta}$ for two particles $\alpha, \beta \in
\{q,b,m\}$, which are typically weak and will be neglected in our treatment. This single-band model is valid
in the limit for $U_{\alpha\beta},J,\Omega,\Delta \ll \omega$, where $\omega$ is the energy separation
between Bloch bands, an inequality which is fulfilled in current experiments. The robustness of the SAT with
respect to loss processes is discussed in \cite{sat}.

In the rest of this article, we will study the current of atoms past the impurity site that develops as a
function of time, and how this current depends on the interaction between probe atoms and on interactions
between the probe atoms and the impurity.

\subsection{Atomic Currents through the SAT}

To analyse the case of many atoms passing the impurity site, we consider the probe atoms $b$ to be prepared
initially to the left of the impurity, in a ground state corresponding to a 1D box potential. The current of
atoms passing the impurity is $I(t)=d{\hat N}_R/dt$, where $N_R$ is the mean number of atoms to the right of
the impurity, $N_R = \langle\sum_{j>0}{\hat b}_j^\dag {\hat b}_j\rangle$. For a sufficiently large number of
atoms in the initial cloud, this current is generally found to rapidly settle into an initial steady state
current, $I_{\rm ss}$, on relatively short timescales ($tJ\sim 1$) (see section \ref{sec:sscurrchi} for
further discussion of steady state currents for bosons).

For a non-interacting Fermi gas at zero temperature, the currents can be calculated exactly when $U_{bm}=
U_{qb}$, as the equations of motion are linear. Scattering from the impurity occurs independently for each
particle in the initial Fermi sea, and after a short transient period of the order of the inverse tunnelling
rate $1/J$, a steady state current $I_{\rm ss}$ is established. This can be calculated either by integrating
the single-particle transmission probabilities \cite{sat,satandi} or by direct numerical integration of the
Heisenberg equations.

For a non-interacting and very dilute Bose gas, the situation will be identical to considering a single
particle. However, for higher densities, many-boson effects become important, and additionally for non-zero
interactions the situation becomes even more complicated. In the limit $U_{bb}/J \rightarrow \infty$ in 1D
(the Tonks gas regime) it is usually possible to replace the bosonic operators $\hat{b}_i, \hat{b}_i^\dag$ by
fermionic operators $\hat{f}_i, \hat{f}_i^\dag$ using a Jordan-Wigner transformation \cite{jordanwigner}.
However, in this case the resulting Hamiltonian,
\begin{eqnarray}
\hat{H}&=&-J\sum_{\langle ij \rangle} {\hat f}_i^\dag {\hat f}_j -\Delta {\hat m}^\dag {\hat
m}+(-1)^{\hat{N}_L}\Omega \left({\hat m}^\dag {\hat q} {\hat f}_0 + {\rm h.c}\right) \nonumber\\
& &+U_{qb} {\hat f}_0^\dag {\hat q}^\dag {\hat q} {\hat f}_0 + U_{bm} {\hat f}_0^\dag {\hat m}^\dag {\hat m}
{\hat f}_0, \label{eq:jwtranform}
\end{eqnarray}
contains a nonlinear phase factor resulting from the coupling on the impurity site, $(-1)^{\hat{N}_L}$, where
$\hat{N}_L=\sum_{j<0}{\hat f}_j^\dag {\hat f}_j$ is the operator for the number of atoms to the left of the
impurity site. For $\Omega=0$, the boson currents are exactly the same as the currents for noninteracting
fermions as $\langle\hat{b}_i^\dag \hat{b}_i\rangle=\langle\hat{f}_i^\dag \hat{f}_i\rangle$. For finite
$\Omega$ it is not clear what role the phase factor will play in determining the system dynamics, although
for sufficiently large $\Omega \gg J$ we again expect very little current to pass the impurity.

Thus, for the intermediate regime $\Omega \sim J$, and for the case of finite interaction strength $U/J$
there are no known analytical solutions for the currents. For this reason, we specifically study these
regimes in this paper, using near-exact numerical methods.

\subsection{Time-Dependent Numerical Algorithm for 1D Many-Body Systems}
\label{sec:overviewtebd}

The algorithm that we use to compute the time evolution of our many body system for bosonic probe atoms was
originally proposed by Vidal \cite{vidal}. This method allows near-exact integration of the many body
Schr\"{o}dinger equation in 1D by an adaptive decimation of the Hilbert space, provided that the Hamiltonian
couples nearest-neighbour sites only and that the resulting states are only ``slightly entangled'' (this will
be explained in more detail below). Recently both this algorithm \cite{dissipativevidal}, and similar methods
proposed by Verstrate and Cirac \cite{dissipativecirac} have been generalised to the treatment of master
equations for dissipative systems and systems at finite temperature, and progress has been made applying the
latter method to 2D systems \cite{cirac2d}.

In 1D, these methods rely on a decomposition of the many-body wavefunction into a matrix product
representation of the type used in Density Matrix Renormalisation Group (DMRG) calculations
\cite{dmrgreview}, which had previously been widely applied to find the ground state in 1D systems. The time
dependent algorithms have now been incorporated within DMRG codes \cite{dmrgvidal}, and also been used to
study the coherent dynamics of a variety of systems \cite{vidalexamples}. In our case, we write the
coefficients of the wavefunction expanded in terms of local Hilbert spaces of dimension $S$,
\beq
\ket{\Psi}=\sum_{i_1 i_2 \ldots i_M = 1}^S c_{i_1 i_2 \ldots i_M} \ket{i_1} \otimes \ket{i_2} \otimes \ldots
\otimes \ket{i_M},
\eeq
as a product of tensors
\beq
c_{i_1 i_2 \ldots i_M}=\sum_{\alpha_1 \dots \alpha_{M-1}}^\chi \Gamma^{[1] \;
i_1}_{\alpha_1}\lambda^{[1]}_{\alpha_1}\Gamma^{[2] \; i_2}_{\alpha_1
\alpha_2}\lambda^{[2]}_{\alpha_2}\Gamma^{[2] \; i_2}_{\alpha_3 \alpha_4} \ldots \Gamma^{[M] \;
i_M}_{\alpha_{M-1}}.
\eeq
These are chosen so that the tensor $\lambda^{[l]}_\alpha$ specifies the coefficients of the Schmidt
decomposition \cite{nielsonchuang} for the bipartite splitting of the system at site $l$,
\beq
\ket{\psi}=\sum_{\alpha=1}^{\chi_l} \lambda^{[l]}_\alpha \ket{\phi_\alpha^{[1 \ldots
l]}}\ket{\phi_\alpha^{[l+1 \ldots M]}},
\eeq
where $\chi_l$ is the Schmidt rank, and the sum over remaining tensors specify the Schmidt eigenstates,
$\ket{\phi_\alpha^{[1 \ldots l]}}$ and $\ket{\phi_\alpha^{[l+1 \ldots M]}}$. The key to the method is
two-fold. Firstly, for many states corresponding to a low-energy in 1D systems we find that the Schmidt
coefficients $\lambda^{[l]}_\alpha$, ordered in decreasing magnitude, decay rapidly as a function of their
index $\alpha$ (this is what we mean by the state being ``slightly entangled'') \cite{vidal}. Thus the
representation can be truncated at relatively small $\chi$ and still provide an inner product of almost unity
with the exact state of the system $\ket{\Psi}$. Secondly, when an operator acts on the local Hilbert state
of two neighbouring sites, the representation can be efficiently updated by changing the $\Gamma$ tensors
corresponding to those two sites, a number of operations that scales as $\chi^3 S^3$ for sufficiently large
$\chi$ \cite{vidal}. Thus, we represent the state on a systematically truncated Hilbert space, which changes
adaptively as we perform operations on the state.

In order to simulate the time evolution of a state, we perform a Suzuki-Trotter decomposition \cite{trotter}
of the time evolution operator $\exp(-\rmi \hat{H} t)$, which is applied to each pair of sites individually
in small timesteps $\delta t$. Initial states can also be found using an imaginary time evolution, i.e., the
repeated application of the operator $\exp(-\hat{H} \delta t)$, together with renormalisation of the state.

In this paper, results are not only produced using the original algorithm as presented in \cite{vidal}, but
also using an optimised version in which the Schmidt eigenstates are forced to correspond to fixed numbers of
particles. This allows us to make use of the total number conservation in the Hamiltonian to substantially
increase the speed of the code, and also improve the scaling with $\chi$ and $S$. With this number conserving
code we are able to compute results with much higher values of $\chi$, however we also find that for
insufficiently large $\chi$, the results from this code become rapidly unphysical, in contrast to the
original code (see section \ref{sec:sscurrchi}).

In implementations of this method we vary the value of $\chi$ to check that the point at which the
representation is being truncated does not affect the final results. A useful indicator for convergence of
the method is the sum of the Schmidt coefficients discarded in each time step, although in practice the
convergence of calculated quantities (such as the single particle density matrix, $\langle \hat{b}_i^\dag
\hat{b}_j\rangle$) are normally used. This is also discussed further in section \ref{sec:sscurrchi}

For bosons on an optical lattice we must also choose the dimension $S$ of the local Hilbert space, which
corresponds to one more than the maximum number of atoms allowed on one lattice site. For simulation of the
SAT, we allow a variable dimension of the local Hilbert space $S_l$, as we must consider the state of the
molecule on the impurity site in addition to the probe atoms. Allowing such a variable dimension dramatically
reduces the simulation time, which scales as $\chi^3 \sum_l S_l^3$ when $\chi \gg S$, and scales proportional
to $S^4$ when $\chi$ is small. For a Bose gas with finite $U/J$ we usually take $S_l=6$ away from the
impurity site, and $S_0=12$ on the impurity site, whereas simulations of a Tonks gas can be performed with
$S_l=2$ away from the impurity site and $S_0=4$ on the impurity site.

\section{Numerical Results}
\label{sec:res}

In section \ref{sec:sscurrchi} we discuss the time dependence of the current for bosons and the applicability
of our numerical methods in different regimes. We establish the existence of an initial steady state current,
$I_{SS}$ that appears on a timescale $tJ\sim 1$, and discuss the observation of a second steady state current
$I_0$, observed in some cases on a timescale $tJ\sim10$. In sections \ref{sec:resdiff} and
\ref{sec:reskicked} we then present our numerical results for $I_{SS}$ for the case where the initial cloud
diffuses past the impurity site, and the case where the initial cloud is kicked respectively.

We are primarily interested in the behaviour of the current through the SAT when it is used in the ``off''
state, i.e., we choose $\Delta=0$. To enhance clarity of the results, we also choose $U_{bq}=U_{bm}=0$.

In each case, we considered an initial cloud of between $N=1$ and $N=30$ atoms, confined on $M=30$ lattice
sites situated immediately to the left of the impurity site. The initial state used corresponds to the ground
state, $\ket{\phi_0}$ of a Bose-Hubbard model with a box trap.

Our total grid for the time evolution consisted of 61 lattice sites, with the 30 rightmost sites initially
unoccupied, and the results we present are, except for very small systems, independent of the size of the
initial cloud and of the grid size. Fermionic results are derived from exact integration of the Heisenberg
equations of motion, whereas bosonic results are near-exact simulations as described in section
\ref{sec:overviewtebd}.

\subsection{Time Dependence of the current for bosonic probe atoms}
\label{sec:sscurrchi}

\begin{figure}[tb]
\begin{center}
\includegraphics[width=\halffigwidth]{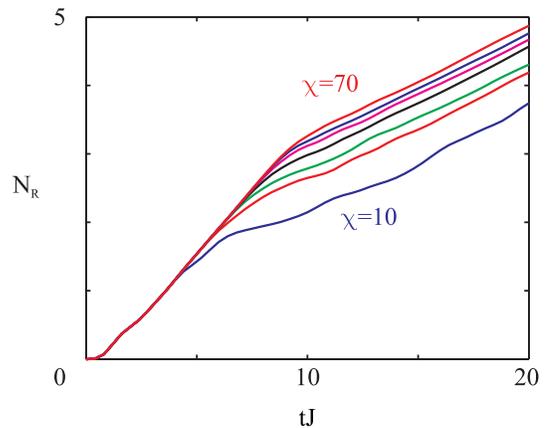}
\caption{The number of atoms to the right of the impurity site, $N_R$ as a function of dimensionless time $t
J$ for a Bose gas in the Tonks limit ($U_{bb}/J\rightarrow \infty$) with $\Omega/J=1$, $n=N/M=1$, and varying
number of states retained in the method, $\chi=10,20,30,40,50,60,70$ (lines from bottom to top). These
results are from the original simulation method.}\label{fig:nr-chiorig}
\end{center}
\end{figure}

The mean number of probe atoms on the right of the impurity, $N_R$ is plotted as a function of time, $t$, in
Fig.~\ref{fig:nr-chiorig} for a Tonks gas ($U/J\rightarrow \infty$) with $\Omega/J=1$ and initial state of
density $n=N/M=1$. These results were calculated with the original simulation algorithm, and it is clear from
the figure that the current settles into an initial steady state value $I_{SS}$ on the timescale $tJ\sim 1$.
However, as is typical for bosonic probe atoms with $n>0.5$, there exists a knee in the curve at a time
$t_{\rm knee}(\chi)$, leading to a new and final steady state current, which we will denote $I_0$. The time
$t_{\rm knee}(\chi)$ depends on the initial density, $n$, and coupling, $\Omega$, and as can be seen from
this figure, we require a high value of $\chi$ to find the exact time. For $n=1,\Omega/J=1$, $t_{\rm
knee}(\chi)$ appears to converge to a value between $tJ=9$ and $tJ=12$ as $\chi$ is increased. It is clear
that significant level of correlation, or entanglement between the left and right hand side of the system (in
the sense of the number of significant Schmidt eigenvalues for a bipartite splitting) are involved in
determining the dynamics leading the to knee. However, the actual value of the steady state current $I_0$
appears to converge for much lower values of $\chi$ and there is essentially no change in this result from
$\chi=10$ to $\chi=70$.

\begin{figure}[ptb]
\begin{center}
\includegraphics[width=\halffigwidth]{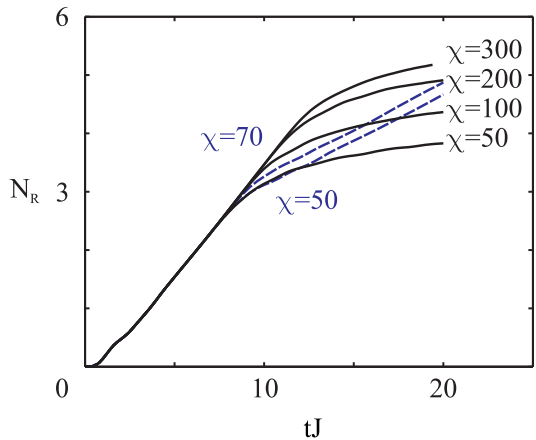}
\caption{The number of atoms to the right of the impurity site, $N_R$ as a function of dimensionless time $t
J$ for a Bose gas in the Tonks limit ($U_{bb}/J\rightarrow \infty$) with $\Omega/J=1$, $n=N/M=1$. This plot
shows a comparison of results from the original method (dashed lines, $\chi=50,70$, c.f.
Fig.~\protect{\ref{fig:nr-chiorig}}), and from the number conserving method (solid lines,
$\chi=50,100,200,300$).}\label{fig:nr-chiopt}
\end{center}
\end{figure}

The interpretation of these results is more complex when they are compared with similar results from the new,
number conserving version of our code. In Fig.~\ref{fig:nr-chiopt} we observe that the behaviour diverges at
the same value of $t_{\rm knee}(\chi)$, and even for $\chi=300$, the value of $t_{\rm knee}(\chi)$ has only
shifted a little further from where it was observed for $\chi=70$ with the original version of the code. This
confirms that the dynamics on this timescale are dominated by the significant level of correlation, or
entanglement between the left and right hand side of the system.

\begin{figure}[ptb]
\begin{center}
\includegraphics[width=\halffigwidth]{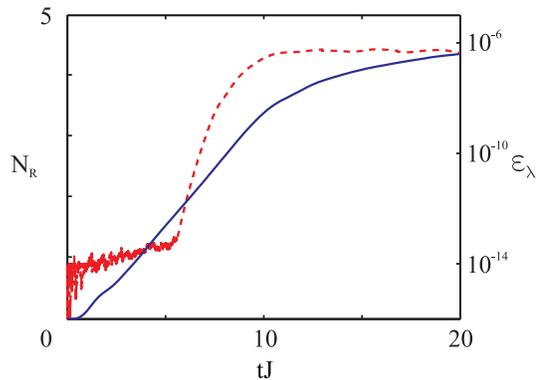}
\caption{Comparison as a function of dimensionless time $t J$ of the number of atoms to the right of the
impurity site, $N_R$, (solid line), and the sum of squares of the discarded Schmidt eigenvalues,
$\varepsilon_\lambda=\sum_{\beta>\chi} \lambda_\beta^2$ (dashed line). These results are taken from the
number conserving simulation method with $\chi=100$, for a Bose gas in the Tonks limit ($U_{bb}/J\rightarrow
\infty$) with $\Omega/J=1$, $n=N/M=1$.}\label{fig:nrerror}
\end{center}
\end{figure}

In contrast to the steady state current $I_0$ obtained using the original code, though, the current in the
number conserving simulations rapidly approaches $0$, even for $\chi=300$. As can be seen from the dotted
line in Fig.~\ref{fig:nrerror}, this behaviour occurs when the maximum sum of squares of the Schmidt
coefficients being discarded in each timestep, $\varepsilon_\lambda=\sum_{\beta>\chi} \lambda_\beta^2$,
reaches a steady value on the order of $10^{-7}$, indicating that the simulation results from the number
conserving code are probably not valid for $t>t_{\rm knee}$. Indeed, we observe the same behaviour from the
new simulation code with $\Omega=0$, where we know from Eq. \ref{eq:jwtranform} that the time dependent
current $I(t)$ is equal to that for fermions, and should not decrease in this manner (see currents for
fermions in Ref. \cite{sat}). Interestingly, the original code, which produces the steady state currents
$I_0$ at finite $\Omega$ reproduces the known result at $\Omega=0$ exactly even for small values of $\chi$,
with a steady state current $I_{SS}$ and no knee.

Our conclusions from these results are as follows:

(i) We know that up to $t_{\rm knee}$ our simulation results are exact, as they are unchanged in the linear
region with current $I_{SS}$ for $\chi=20\rightarrow 300$. As this regime lasts at least until $tJ\sim10$,
these results would be observable in an experimental implementation of the SAT.

(ii) As an impractically large value of $\chi$ would be required to reproduce the results exactly on long
timescales, we can not be certain what the final behaviour will be for $t>t_{\rm knee} (\chi=300)$. This
depends on clearly interesting phenomena that arise from strong correlations between the left and right sides
of the impurity site, and could include settling to a final steady state current $I_0$. These effects would
also be observable in an experiment.

The expected final steady state values $I_0$ are already discussed in Ref. \cite{sat}, and so in the
remainder of this article we investigate the initial steady state currents $I_{SS}$ in various parameter
regimes.

\subsection{Diffusive evolution, with initial mean momentum $(\langle \hat{k} \rangle_{t=0}=0)$}
\label{sec:resdiff}

We first consider the motion of atoms past the impurity site in the diffusive regime, where the initial state
at $t=0$ is the ground state of a Bose-Hubbard model on $M=30$ lattice sites in a box trap.

\subsubsection{Dependence of the current on impurity-probe coupling, $\Omega$}

\begin{figure}[ptb]
\begin{center}\includegraphics[width=8cm]{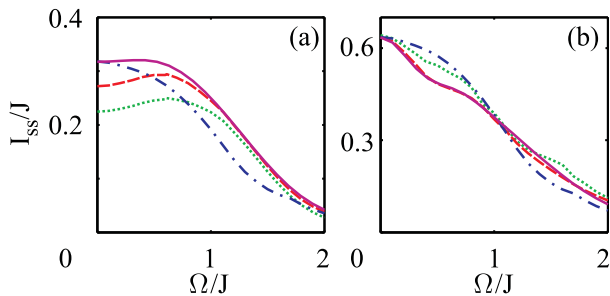}
\caption{Steady state currents through the SAT $I_{SS}$ as a function of the coupling between probe atoms and
the impurity, $\Omega/J$. These plots show the comparison of a Bose gas with different interaction strengths
$U/J=4$ (dotted line), $U/J=10$ (dashed) and $U/J\rightarrow\infty$ (solid), and a Fermi gas (dash-dot), with
(a) $n=1/2$ and (b) $n=1$. In both cases, $\Delta=U_{qb}=U_{bm}=0$. } \label{fig:om-q0}
\end{center}
\end{figure}

In Fig.~\ref{fig:om-q0} we show the initial steady state current $I_{SS}$ as a function of $\Omega/J$ for
fermionic probe atoms, and for bosonic probe atoms with $U_{bb}/J=4,10,\infty$ and $\Delta=0$. All of these
results decrease as expected with increasing $\Omega/J$, and even for a relatively small $\Omega=2J$ the
current is minimal in each case. At half filling (Fig.~\ref{fig:om-q0}a), the results for the Tonks gas are
identical to the Fermi results for $\Omega=0$, but become substantially different as $\Omega$ increases, with
the currents in this regime greater for the bosons. At weaker interactions the currents are smaller than the
Tonks result at all $\Omega$, but for $\Omega/J>1$ the currents for $U/J=4$ are larger than for a
non-interacting Fermi gas. The variation in the currents for different interaction strengths of bosons
appears to be due to the broader initial momentum distributions that occur at larger $U/J$. At unit filling
(Fig.~\ref{fig:om-q0}b), $I_{SS}$ is less dependent on the interaction strength, with all of the bosonic
results very close to one another, currents becoming larger than that for fermions when $\Omega/J>1$.

\subsubsection{Dependence of the current on interaction strength, $U/J$}

\begin{figure}[ptb]
\begin{center}\includegraphics[width=8cm]{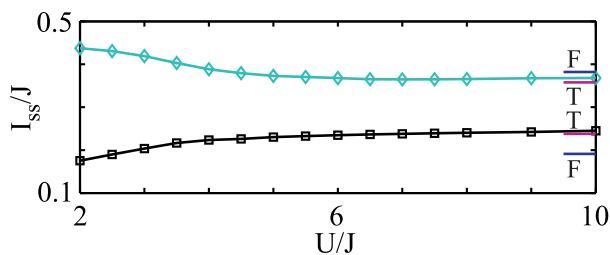}
\caption{Steady state currents through the SAT $I_{ss}$ as a function of the interaction strength $U/J$ for
bosonic probe atoms initially at half-filling, $n=1/2$ (squares) and at unit filling, $n=1$ (diamonds) with
$\Omega/J=1$. The equivalent results for the Tonks gas ($U/J\rightarrow \infty$) and fermions are marked on
the right hand site of the plot. $\Delta=U_{qb}=U_{bm}=0$.} \label{fig:uj-q0}
\end{center}
\end{figure}

The dependence of the initial steady state current $I_{SS}$ on the interaction strength for bosons is
depicted more clearly in Fig.~\ref{fig:uj-q0}, both at unit filling, $n=1$, and half filling, $n=1/2$ for
$\Omega=J$. At half filling the current increases with increasing $U/J$, which is due to the broader initial
momentum distribution produced by the higher interaction energies. In contrast, at higher densities (here
$n=1$), the probe atoms are blocked better by the SAT for higher interaction strengths, and $I_{SS}$
decreases. The key principle here is that bosons appear to be better blocked when they approach the impurity
individually. For high densities this is achieved when large interaction strengths eliminate the higher
occupancies of all lattice site including the impurity site. For weaker interactions the bosons can swamp the
transistor, with one atom being bound to the impurity, whilst other probe atoms tunnel onto and past the
impurity site.

\begin{figure}[ptb]
\begin{center}\includegraphics[width=8cm]{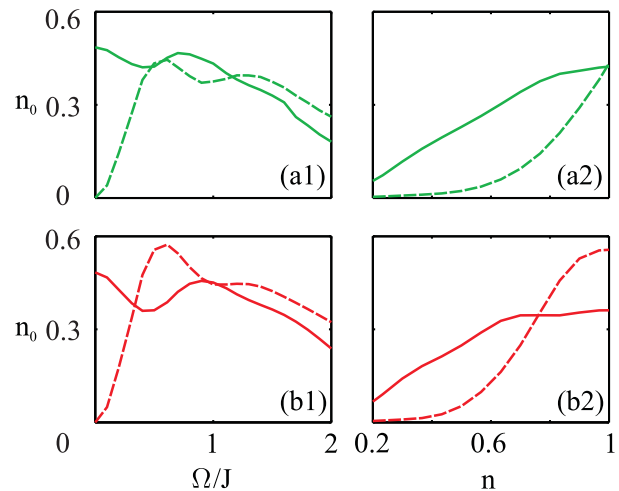}
\caption{Plot showing the average steady state occupation of the molecular state (dashed lines) and the
average steady state atomic occupation of the impurity site (solid lines) for (a) $U/J=4$ and (b) $U/J=10$,
as a function of ($a,b$ 1) $\Omega/J$ with $n=1$ and ($a,b$ 2) $n$ with $\Omega/J=0.5$. In all cases,
$\Delta=U_{qb}=U_{bm}=0$, and calculations were performed for $M=30$.} \label{fig:molocc}
\end{center}
\end{figure}

This effect is seen in Fig.~\ref{fig:molocc}, where the molecular occupation and average probe atom
occupation on the impurity site are shown for (a) $U/J=4$ and (b) $U/J=10$. We see that as $n$ increases, the
molecular occupation becomes rapidly higher for $U/J=10$ than for $U/J=4$, despite the larger occupation of
probe atoms on the impurity site for $U/J=4$. This indicates that for $U/J=10$ atoms arrive individually at
the impurity site, where they are coupled with the impurity atom into a molecular state, and their transport
is efficiently blocked. For $U/J=4$, more than one atom enters the impurity site at once, leading to a larger
average probe atom occupation on the impurity site, but a comparatively small molecular occupation.

It is important to note, however, that even when $U/J=4$, the resulting currents are only slightly larger
than they are for non-interacting fermions. At higher interaction strengths we then see an even stronger
suppression of the steady state current for dense, strongly interacting bosons. As $\Omega$ increases, both
the molecular occupation and probe atom occupation on the impurity site decrease (Fig.~\ref{fig:molocc}) as
the probability of even a single atom tunnelling onto the impurity site becomes small. For $\Omega>2J$ the
blocking mechanism of the SAT functions extremely well even in the regime where the probe atoms are dense and
weakly interacting.

\subsubsection{Dependence of the current on initial density, $n$}

\begin{figure}
\begin{center}
\includegraphics[width=8cm]{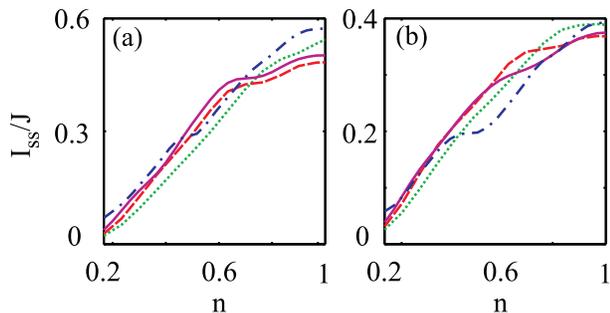}
\caption{Steady state currents through the SAT $I_{ss}$ as a function of the initial density of atoms
$n=N/M$. These plots show the comparison of a Bose gas with different interaction strengths $U/J=4$ (dotted
line), $U/J=10$ (dashed) and $U/J\rightarrow\infty$ (solid), and a Fermi gas (dash-dot), with (a)
$\Omega/J=0.5$ and (b) $\Omega/J=1$. In both cases, $\Delta=U_{qb}=U_{bm}=0$.} \label{fig:n-plot}
\end{center}
\end{figure}

In Fig.~\ref{fig:n-plot} we show the dependence of the initial steady state current, $I_{SS}$ on the initial
filling factor $n$ with (a) $\Omega/J=0.5$ and (b) $\Omega/J=1$. In both cases, the currents for bosons of
different interaction strengths are very similar, with the variations following the patterns discussed in the
preceding section. These results also agree well with the results for fermions at small $n$ and for $n\sim
1$, but the plateau observed in fermionic currents near $n\sim 0.5$ does not occur in the currents for
bosons. For fermions, this plateau arises from the transmission profile of the SAT as a function of incoming
momentum \cite{sat}, and occurs when the Fermi momentum is raised past the minimum in this transmission
profile. For interacting bosons, this correspondence between the momentum distribution of the gas and the
transmission profile is destroyed by many-body effects, and we see instead a smooth increase in the current.
This results in the bosonic currents being substantially larger than those for fermions near half filling
when $\Omega \sim 1$ (as was previously observed in Fig.~\ref{fig:om-q0}a).

\subsection{Kicked evolution, with initial mean momentum $(\langle \hat{k} \rangle_{t=0} \neq 0)$}
\label{sec:reskicked}

In this section we consider an initial state with a non-zero initial momentum, which is obtained, e.g., by
briefly tilting the lattice on a timescale much shorter than that corresponding to dynamics of atoms in the
lattice. If the tilt is linear, the resulting state will be given by
\beq
\ket{\phi(t=0)}=\sum_j\exp (\rmi p_k j \hat{b}_j^\dag\hat{b}_j) \ket{\phi_0},
\eeq
where $\ket{\phi_0}$ is the initial many-body ground state, and the quantity $p_k$ is determined by the
magnitude and duration of the tilt. The effect of this tilt is to translate the ground state in the periodic
quasimomentum space by a momentum $p_k$. The final mean momentum $\langle k \rangle$ then depends both on the
value $p_k$ and the properties of the initial momentum distribution.



\subsubsection{Dependence of the current on kick strength $p_k$}

\begin{figure}
\begin{center}
\includegraphics[width=8cm]{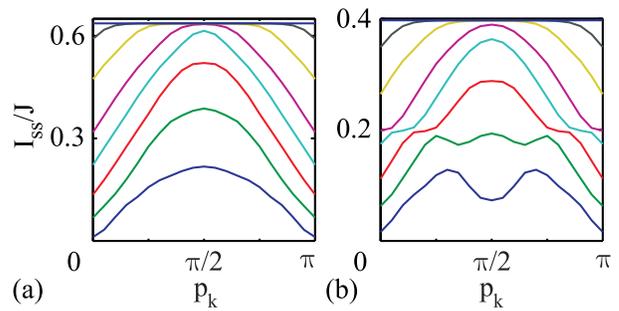}
\caption{Steady state currents $I_{SS}$ through the SAT for fermions as a function of the kick parameter
$p_k$, for (a) $\Omega=0$ and (b) $\Omega/J=1$.  In each plot the lines from bottom to top sequentially
correspond to filling of $N=3,6,9,12,15,20,25,30$ particles initially on $M=30$ lattice sites. Note that the
scales are different for (a) and (b), and also that the results in (a) are exactly the same as those for a
Tonks gas of bosons. In both cases, $\Delta=U_{qb}=U_{bm}=0$.} \label{fig:fermikick}
\end{center}
\end{figure}

In the case of fermions, the dependence of the current on $q$ for different filling factors $n=N/M$ and
$\Omega$ can be clearly understood in terms of the SAT transmission profile (see \cite{satandi}). In
Fig.~\ref{fig:fermikick}a we see the current $I_{ss}$ as a function of $p_k$ with $\Omega=0$. The currents
are each peaked at $p_k=\pi/2$, where the resulting mean velocity of the probe atoms is the largest. For
$N/M=1$, the whole Bloch band is filled, and the momentum distribution is not changed by the application of
the kick, i.e., $\langle \hat{k}\rangle_{t=0}=0$. In Fig.~\ref{fig:fermikick}b the same results are shown,
but with $\Omega/J=1$. Here we see that for small filling factors, a minimum appears at $p_k=\pi/2$,
corresponding to the minimum in the transmission profile of the SAT for this incident momentum
\cite{sat,satandi}. At higher filling factors, this feature of the transmission profile for $\Omega/J=1$ is
not sufficiently broad to overcome the increase current due to higer mean velocities in the initial cloud,
and the peak at $p_k=\pi/2$ reappears. The currents here are, of course, reduced in comparison with those for
$\Omega=0$.

\begin{figure}[ptb]
\begin{center}\includegraphics[width=8cm]{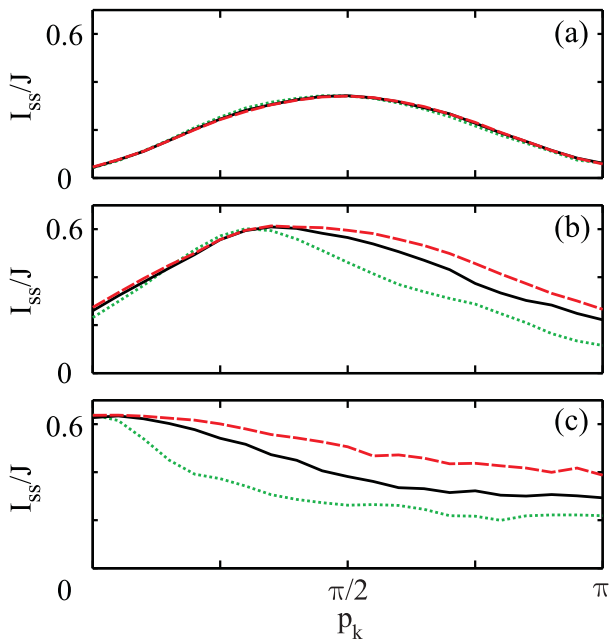}
\caption{Steady state currents with coupling to the SAT, $\Omega=0$, $I_{SS}$ as a function of the kick
parameter $p_k$ for varying interaction strengths, $U/J=4$ (dotted), $U/J=7$ (solid), and $U/J=10$ (dashed),
for (a) $N=5$, (b) $N=15$ and (c) $N=30$ particles initially situated on $M=30$ lattice sites. In all cases,
$\Delta=U_{qb}=U_{bm}=0$.} \label{fig:q-om0}
\end{center}
\end{figure}

Whilst for all $p_k$ the currents with no coupling to the impurity atom, i.e., $\Omega=0$, are the same for
the Tonks gas as for fermions (Fig.~\ref{fig:fermikick}a), the currents for finite interaction strengths are
found to be remarkably different. In Fig.~\ref{fig:q-om0} these rates are plotted for $U/J=4,7,10$ for
$N=5,15,30$ particles initially situated on $M=30$ sites. For the very dilute system with $N=5$
(Fig.~\ref{fig:q-om0}a) we see a peak similar to that observed for fermions which is independent of the
interaction strength. Here the currents are essentially those for non-interacting particles, and the currents
determined by the initial momentum distribution. For $N=15$ (Fig.~\ref{fig:q-om0}b) we observe the surprising
result that the current is peaked at a lower value than is observed for fermions or for the Tonks gas. We
have observed this peak consistently for such cases of finite interaction, and note that as $U/J$ increases,
the peak moves back towards $p_k=\pi/2$ as the currents converge to the Tonks gas results. As $N$ is further
increased, the peak continues to move left, and for $N=30$ (Fig.~\ref{fig:q-om0}c) we see a monotonically
decreasing current as $p_k$ increases. As $U/J$ increases these values tend towards the $p_k$ independent
result observed for the Tonks gas. These results are surprising, but the trends in the behaviour are clear,
and they should be directly verifiable in experiments, even without the presence of the impurity atom.

\begin{figure}[ptb]
\begin{center}\includegraphics[width=8cm]{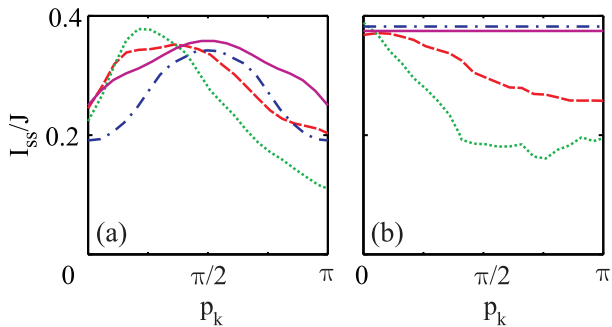}
\caption{Steady state currents through the SAT $I_{SS}$ as a function of the kick parameter $p_k$, for
$\Omega/J=1$. These plots show the comparison of a Bose gas with different interaction strengths $U/J=4$
(dotted line), $U/J=10$ (dashed) and $U/J\rightarrow\infty$ (solid), and a Fermi gas (dash-dot), with (a)
$n=1/2$ and (b) $n=1$. In both cases, $\Delta=U_{qb}=U_{bm}=0$.} \label{fig:q-om1}
\end{center}
\end{figure}

For non-zero coupling to the impurity atom, the currents as a function of $p_k$ are shown in
Fig.~\ref{fig:q-om1}. Again we notice that the current for bosons with finite interaction strength is peaked
at much lower values of $p_k$ than the fermionic currents and that peaks of all of the bosonic currents,
including the Tonks currents, as significantly larger than the fermionic currents at half filling, as was
observed for diffusive results ($p_k=0$). The most remarkable feature of these plots is that despite a
significant reduction in the current, the basic dependence on $p_k$ is very similar to the $\Omega=0$
results.

\subsubsection{Dependence of the current on impurity-probe coupling, $\Omega$}

\begin{figure}[tb]
\begin{center}\includegraphics[width=8cm]{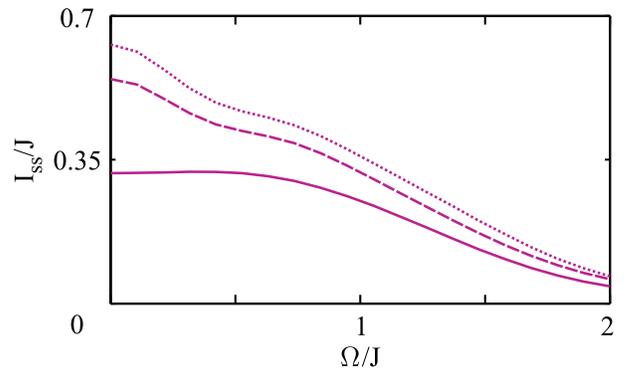}
\caption{A comparison of steady state currents through the SAT, $I_{SS}$, as a function of $\Omega/J$ for
$p_k=0$ (solid line), $\pi/4$ (dashed), and $\pi/2$ (dotted). Here we consider a Tonks gas
($U/J\rightarrow\infty$) of bosonic probe atoms which is initially at half half-filling, $N=15$, $M=30$.
$\Delta=U_{qb}=U_{bm}=0$.} \label{fig:om-q-tonks}
\end{center}
\end{figure}

The steady state current $I_{ss}$ is shown in Fig.~\ref{fig:om-q-tonks} as a function of $\Omega$. We observe
the same strong decrease in the current due to the operation of the SAT for all of these curves, with the
highest currents corresponding to the $p_k=\pi/2$ curve as expected. Note that the value of $I_{SS}$ is
affected equally for all of the kick strengths, and the ratio in the currents for different values of $p_k$
is very similar for $\Omega=0$ and $\Omega/J=2$.

\section{Summary}
\label{sec:summary}

In summary, the SAT setup provides new experimental opportunities to study coherent transport of many atoms
past a spin-1/2 impurity due to the relatively long coherence times that exist for systems of atoms in
optical lattices. The resulting coherent many-body effects can be clearly seen in the difference between the
atomic currents observed for fermions and bosons, and the non-trivial dependence of the current on
interaction strength for bosons with finite interactions. Even stronger dependence on these interactions is
observed when the probe atoms are initially accelerated to a non-zero momentum. The initial steady state
currents would be directly accessible quantities in the experimental implementation of the SAT, and using
recently developed methods for time-dependent calculation of many-body 1D systems, we have made quantitative
predictions for the corresponding currents for a wide range of system parameters. We cannot be certain about
the values the currents approach at long times, although it is possible that the system will settle
eventually into a regime with a different steady state current. The high values of $\chi$ needed to reproduce
this behaviour in our numerical calculations suggest that the currents in this regime could also be strongly
sensitive to the coherence properties of the system, which would be very interesting to investigate in an
experiment.

\begin{acknowledgments} The authors would like to thank A. Micheli for advice and stimulating discussions, G. Vidal for
helpful discussions on the numerical methods, and A. Kantian for his contributions to producing the number
conserving version of the program code. AJD thanks the Clarendon Laboratory and DJ thanks the Institute for
Quantum Optics and Quantum Information of the Austrian Academy of Sciences for hospitality during the
development of this work. This work was supported by EU Networks and OLAQUI. In addition, work in Innsbruck
is supported by the Austrian Science Foundation and the Institute for Quantum Information, and work in Oxford
is supported by EPSRC through the QIP IRC (www.qipirc.org) (GR/S82176/01) and the project EP/C51933/1.
\end{acknowledgments}

\end{document}